\documentclass[11pt,a4paper]{article}
\pdfoutput=1
\usepackage{jheppubnohead}
\usepackage{tabularx}
\usepackage[normalem]{ulem} 
 \usepackage{amsmath,amsfonts,amssymb,dsfont,mathrsfs,graphicx, lieart, booktabs, siunitx, caption, subcaption,esint}
\usepackage[capitalise, english]{cleveref}

\newcommand{\eq}[1]{Eq.~(\ref{#1})}
\newcommand{\bea}{\begin{eqnarray}}
\newcommand{\eea}{\end{eqnarray}}

\newcommand{\beq}{\begin{equation}}
\newcommand{\eeq}{\end{equation}}
\newcommand{\be}{\begin{equation}}
\newcommand{\ee}{\end{equation}}
\newcommand{\beqn}{\begin{eqnarray}}
\newcommand{\eeqn}{\end{eqnarray}}

 \title{\begin{center}Is the Relaxion an Axion?\end{center}}
 
 \author[a]{Rick~S. Gupta,}
 \author[a]{Zohar Komargodski,}
  \author[a]{Gilad Perez,}
 \author[b]{Lorenzo Ubaldi}

\affiliation[a]{Department of Particle Physics and Astrophysics, \\
Weizmann Institute of Science, Rehovot 76100, Israel}

\affiliation[b]{Raymond and Beverly Sackler School of Physics and Astronomy, \\
 Tel-Aviv University, Tel-Aviv 69978, Israel}

\emailAdd{rsgupta@weizmann.ac.il}
\emailAdd{zohar.komargodski@weizmann.ac.il}
\emailAdd{gilad.perez@weizmann.ac.il}
\emailAdd{ubaldi.physics@gmail.com}

\abstract{We consider the recently proposed cosmological relaxation mechanism where the hierarchy problem is ameliorated, and the electroweak (EW) scale is dynamically selected by a slowly rolling axion field. We argue that, in its simplest form, the construction breaks a gauge symmetry that always exists for pseudo-Nambu-Goldstone bosons (in particular the axion). The small parameter in the relaxion model is therefore not technically natural as it breaks a gauge symmetry rather than global symmetries only.  The consistency of the theory generically implies that the cutoff must lie around the electroweak scale, but not qualitatively higher. 
We discuss several ways to evade the above conclusion. Some of them may be sufficient to increase the cutoff to the few-TeV range (and therefore may be relevant for the little-hierarchy problem). To demonstrate the ideas in a concrete setting we consider a model with a familon, the Nambu-Goldstone boson of a spontaneously broken chiral flavor symmetry. The model has some interesting collider-physics aspects and contains a viable weakly interacting dark matter candidate.}

\begin{document}

\maketitle


\section{Introduction and Review} \label{sec:Intro}

The authors of Ref.~\cite{Graham:2015cka} have recently proposed a framework to address the hierarchy problem. The model promotes the squared Higgs mass parameter ($\mu^2$ in the equations below) to a dynamical variable that evolves during inflation and finally stabilizes at a small negative value. As a result, the physical mass of the Higgs  is much smaller than the cutoff of the theory, solving the hierarchy problem (see also Refs.~\cite{Dvali:2003br, Espinosa:2015eda,Hardy:2015laa,Patil:2015oxa,Antipin:2015jia,Jaeckel:2015txa, Kilic:2015joa} for further work on this subject).

We briefly review the first model of Ref.~\cite{Graham:2015cka}, which is the simplest one in terms of field content.\footnote{Note that this  model is ruled out because it predicts an ${\cal O}(1)$ value for $\theta_{\rm QCD}$. More precisely, one could always add a bare $\theta_{\rm QCD}$ and tune the theta angle, but we would like to avoid fine-tuning. This shortcoming will not affect the main points we want to make.} The value of $\mu^2$, the  mass-squared term in the Higgs potential,
\be \label{Higgspotential}
V(h)=\mu^2 h^\dagger h+\lambda (h^\dagger h)^2  \, ,
\ee
changes during the course of inflation as it varies with the classical value of a scalar field $\phi$,
\be \label{mu2}
\mu^2=g \Lambda \phi-\Lambda^2 \, ,
\ee 
which slowly rolls because of a potential
\be \label{Vphi}
V(\phi)=g \Lambda^3 \phi +g^2 \Lambda^2 \frac{\phi^2}{2} + \cdots\,.
\ee
In these equations $\Lambda$ is the cutoff scale of the theory, and the coupling constant $g$ is dimensionless (it is related to $g_{\rm GKR}$, the one in Ref.~\cite{Graham:2015cka}, through $g_{\rm GKR} = \Lambda g$).
During inflation the field $\phi$ slowly rolls from the initial large field value $\phi > \Lambda / g$, such that $\mu^2$ is positive and the electroweak symmetry unbroken, down the $V(\phi)$ potential. It stops rolling shortly after the point $\phi \sim \Lambda / g$ where $\mu^2<0$ because, as the Higgs field obtains a vacuum expectation value (VEV), $v = 174\,$GeV, a feedback potential for $\phi$ is induced via a mechanism that we review below. 
The challenge is to explain why $v \sim |\mu| \ll \Lambda$. 

In the first proposal of Ref.~\cite{Graham:2015cka}, $\phi$  is the QCD axion with a decay constant $f_a$.
The axion couples to gluons via the term $\frac{g^2}{32 \pi^2}\frac{\phi}{f_a} G \wedge G$, with $G$ the gluon field strength. Non-perturbative effects induce a potential for $\phi$ below the confinement scale. The potential can be written as
\be
 \Delta V \sim y_u v f_\pi^3 \cos \left(\frac{\phi}{f_a}\right) \, .
 \label{axpot}
\ee
Here $f_\pi^3 \sim  \langle q \bar{q} \rangle$ is the pion decay constant, $y_u$ the up-quark yukawa coupling, and for simplicity we only retained the leading contributions. Note that this potential respects a discrete symmetry
\be\phi\rightarrow \phi+2\pi k f_a ~,\qquad k\in \mathbb{Z} \, , \label{discrete}\ee
while the terms in \eq{Vphi} explicitly break it.
The formula~\eqref{axpot} only holds when $y_u v\ll \Lambda_{\rm QCD}$. In this regime the amplitude of these oscillations~\eqref{axpot} grows linearly with $v$. When the maximum slope of $\Delta V$ matches the slope of $V(\phi)$, the field stops rolling. This is achieved when 
\begin{equation} 
g\sim  \frac{y_u v f_\pi^3}{\Lambda^3f_a}\,.
\label{gaxion}
\end{equation}
The fine tuning problem is ameliorated if $\Lambda \gg v$ is allowed by the framework.
This in turn implies $g\ll 1$. For example, if we take $f_a = 10^9\,$GeV and $\Lambda \sim 10^7\,$GeV, then $g \sim 10^{-30}$. 

The authors of Ref.~\cite{Graham:2015cka} claim that such a small $g$ is technically natural,
because in the limit $g \to 0$ one recovers a shift symmetry for $\phi$ (in particular that of~\eq{discrete}) and therefore it satisfies the criterion of 't Hooft.

In the next section we explain why having a small $g$ is not natural (and in fact theoretically inconsistent under the assumptions we state carefully in section~\ref{sec:critics}). The main point is that the discrete symmetry~\eqref{discrete} is necessarily {\it gauged} and hence it cannot be broken by any term in the action. This is true as long as $\phi$ is an axion or any other pseudo-Nambu-Goldstone Boson (pNGB). Therefore, a small nonzero $g$ breaks a gauge symmetry rather than a global symmetry. This statement can be converted into an inequality that the parameters need to satisfy for the consistency of the theory, and there are no solutions with cutoff $\Lambda\gg v$. We discuss possible ways out of this conclusion. 

To demonstrate the point further, in section~\ref{sec:model} we provide a phenomenological model where the axion is replaced by a familon (which is another example of a pNGB). 
If the cutoff is high, the model suffers from the generic problems discussed in section 2. However, we investigate whether one may have a slight parametric boost of the cutoff to a couple of TeV and thus render the model interesting in view of the little hierarchy problem. This model also happens to contain a weakly interacting massive particle (WIMP) with the correct relic abundance.
The main conclusion is that, under some assumptions, the relaxion is not an axion (and also not any other pNGB) with period $2 \pi f$, but there are some possible directions that would be interesting to explore within this framewrok.

\section{Is the relaxion a pseudo-Nambu-Goldstone boson?}
\label{sec:critics}
In this section we describe the main problem that arises when constructing a cosmological relaxation model. 
We first discuss the axion case and then generalize the argument to any pNGB field.
We conclude the section with some speculations on how to evade the problems discussed herein.

The fundamental distinction between gauged and global symmetries at the level of effective Wilsonian actions is that there exist operators that can break any global symmetry but there do not exist operators that break gauge symmetries. (One can also say that the operators that break a gauge symmetry are ill-defined -- they cannot be local operators.)

Let us consider a Nambu-Goldstone field (it can be an axion or any other realization of a Nambu-Goldstone field). If some exact global symmetry of the theory is spontaneously broken then the effective action respects a continuous symmetry
\be\label{contshift}\phi \to \phi +\alpha f_a\,,\ee
with any real $\alpha$. But a subgroup of~\eqref{contshift} is gauged, namely, 
\be\phi\rightarrow \phi+2\pi k f_a \,,\qquad k\in \mathbb{Z} \,, \label{discshift}\ee
is a gauge symmetry. It has to do with the fact that the field $\phi$ is fundamentally an angle and hence no local operator can break its periodicity.
Explicit (e.g. soft) breaking of the original global symmetry can result in breaking~\eqref{contshift} to~\eqref{discshift}, but the latter can never be broken.\footnote{Perhaps to make this discussion less abstract one can think of a $U(1)$ global symmetry that is spontaneously broken at some scale $f$ by the VEV of a complex scalar field $\Phi$. A non-linear mapping $\Phi\to \rho \exp[i\phi/f]$, with $\rho$ the modulus of $\Phi$ and $\phi$ its phase ($\rho,\phi$ being real), clearly implies that the shift $\phi\to \phi+2 \pi k f$ maps the field $\Phi$ onto itself. This discrete shift is a redundancy in the description, i.e. a gauge symmetry. Any operator involving $\Phi$, including those that break the global symmetry, will  respect this discrete  gauge symmetry, simply as a consequence of the way $\phi$ has been defined.} For example, the quark masses in QCD break the continuous shift symmetry~\eqref{contshift} to~\eqref{discshift} (more precisely, to the non-Abelian generalization of~\eqref{discshift}). Another example is the coupling $\sim \int d^4x {\phi\over f} G\wedge G$ (which can be induced by an anomaly in the underlying theory) that potentially breaks the continuous shift symmetry but always preserves~\eqref{discshift}. Note that fundamentally there is no distinction between breaking a global symmetry by an ABJ anomaly versus breaking it via an explicit operator (such as a mass term). In fact, sometimes the two can be related to each other by a duality. So our two examples are really not different from each other.

Let us now discuss the consequences of these well-known observations~\cite{Coleman:1969sm} for the models of the type reviewed in the introduction. 

The potential~\eqref{Vphi}  breaks the gauge symmetry~\eqref{discshift}. So if one would like to think of $\phi$ as a pNGB (in particular, an axion), it cannot come from a local QFT unless one can actually rewrite the potential~\eqref{Vphi} in terms of functions with period $2\pi f_a$. Similarly, the effective Higgs mass~\eqref{mu2} should be $2\pi f_a$ periodic.

Since the model requires a non-periodic field excursion of order $\phi\sim \Lambda/g$, we must have $f_a>\Lambda/g$. 
  Using \eq{gaxion} this translates into 
 \begin{equation}
 \Lambda<  100 {\rm \, GeV} \left( \frac{\Delta m_f (v)}{100 {\rm \,GeV}}\right)^\frac{1}{4} \left({\tilde f_\pi}\over 100\ \rm GeV \right)^{3\over 4}\, .
 \label{Lamax}
 \end{equation}
In this inequality we have allowed the axion to couple to a new strong sector, rather than QCD,  with a dynamical symmetry breaking scale $\tilde{f}_\pi$, and we have taken $\Delta m_f(v)$, the contribution of the Higgs VEV to the mass of a new fermion in the strong sector,  to be of order of the weak scale. (The QCD case is recovered by taking $\Delta m_f(v) \to y_u v\sim$ MeV and $\tilde f_\pi\rightarrow f_\pi \sim$ 100 MeV.)
 This results in a cutoff at the electroweak scale, therefore not solving the hierarchy problem. The QCD example in the introduction results in a sub-GeV cutoff. 

As we now show, the same result holds for any pNGB field, i.e., the cutoff cannot be raised parametrically above the electroweak scale. 
The back-reaction potential for a generic field can have various powers of the Higgs VEV and some
dimensionful breaking parameter, $M$ (see also Ref.~\cite{Espinosa:2015eda} for a discussion),  
\be
\Delta V \sim (y v)^n M^{4-n} \cos\left( \frac{\phi}{f}\right) \, .
\ee
We require that $n\geq 0$, which amounts to saying that the back-reaction only turns on when the Higgs field condenses (a necessary feature of the relaxion scenario). The scale $M$ is left arbitrary at this point. If the back-reacting sector breaks the electroweak symmetry then $M$ obviously cannot exceed the electroweak scale. If the back-reacting sector does not break the electroweak symmetry, then one always has the two-loop effect pointed out in Ref.~\cite{Espinosa:2015eda} (we will discuss this effect below and also in section \ref{sec:model}). This does not destroy the framework only if $M$ is not larger than $4\pi v$.

The rolling of $\phi$ stops if $g\sim  \frac{(y v)^n M^{4-n}}{\Lambda^3 f}$  and $\Delta \phi\sim \Lambda/g\lesssim f$ is required for the locality of the effective field theory, as discussed above. This translates into 
 \begin{equation}
 \frac{\Lambda^4}{M^{4}}< \frac{(yv)^n}{M^n}~.
 \label{Lpmax}
 \end{equation}
Since the right-hand side is generically smaller than one and the left hand side is generically bigger than one, this can only be satisfied if all the scales, including the cutoff, are of the order of the Electroweak scale.  
 
In the following we sketch a few potential ideas that might help evade the no-go ``theorem'' above.
\begin{itemize}
\item[(i)] It could be that the relaxion field $\phi$ is not a Peccei-Quinn axion, or more generally is not a Nambu Goldstone boson. Then it is allowed to be fundamentally non-compact. We then only have the global symmetry~\eqref{contshift} and the discrete gauge symmetry is not imposed. The smallness of $g$ can be technically natural. 
\end{itemize}

First, let us note a simple point: in unitary QFTs all the (linearly realized) global symmetries are compact. Therefore, for an effective field theory to be natural, one must similarly require that all of its approximate global symmetries are compact (if they are to be eventually restored).

If $\phi$ is not a pNGB one faces the challenge of explaining where a non-compact, light scalar field comes from. One known mechanism is supersymmetry. In supersymmetric theories the bottom component of a chiral superfield is a complex scalar field and it is natural that its imaginary part would be a pNGB while the real part is a non-compact scalar field. In fact, in supersymmetric vacua the compact symmetry group $G$ of the theory is complexified~\cite{Wess:1992cp} $G\rightarrow G_{\mathbb{C}}$ and this is why there are generically light non-compact scalar fields in SUSY vacua.\footnote{In fact, in renormalizable O'Raifeartaigh models, also SUSY-breaking vacua possess very light non-compact scalar fields (pseudo-moduli)~\cite{Ray:2006wk,Sun:2008nh,Komargodski:2009jf,Curtin:2012yu}. They might be useful for the relaxion framework. Of course, these models would face the basic tension that the non-compact fields' masses are typically not much smaller than the visible gaugino mass. This can be perhaps addressed by complicating the models (and perhaps avoiding gauge mediation).  Another possibly useful class of SUSY-breaking models with light non-compact fields are variations of~\cite{Affleck:1983mk,Affleck:1984xz,Dumitrescu:2010ca}.} In this case one would not expect periodic potentials for the non-compact component of the chiral superfield, so one would have to search for a new backreaction mechanism, which may not be impossible. Another known construction in QFT that allows for a non-compact scalar field is provided by the dilaton, i.e. the Nambu-Goldstone boson of conformal symmetry breaking. The main obstacle here (other than the dilaton would not couple to $G\wedge G$ but to $G\wedge \star G$) is that, apart from trivial free-field examples, currently we do not know of non-supersymmetric,  finite-N, theories that break the conformal symmetry spontaneously. (Even if such theories existed, to lift the moduli space slightly, one would need to have very-nearly-marginal operators. This again can be thought of as a problem of tuning in the space of theories. In terms of the dual AdS theory, this corresponds to having a very light scalar field.)

Another important point is related to the fact that physics at the Planck scale respects gauge symmetries but not global symmetries (see Ref.~\cite{Banks:2010zn} and references therein). Consider the pNGB scenario first. We might worry about operators of the type 
$$M_{{\rm Pl}}^4\int d^4x \cos(\phi/f)~,$$
respecting the gauge symmetry~\eqref{discshift} but violating the global symmetry~\eqref{contshift}. Such operators are disastrous for the phenomenology of these models. A nice way to rule out these operators is by imagining that the UV is a five-dimensional theory on a circle (i.e. $\mathbb{R}^{3,1}\times \mathbb{S}^1$) and the relaxion $\phi$ is just the holonomy of the five-dimensional gauge field along the circle. Then there are no local counter-terms in five dimensions which induce a potential for the holonomy.\footnote{For various applications of this observation see e.g.~\cite{ArkaniHamed:2003wu,Hosotani:1983xw,Rattazzi:2000hs,Feng:2003mk,Cacciapaglia:2007fw,Perez:2008ee}.  This fact has also played a crucial role in compactifications of Yang-Mills theory on a circle (be it a thermal circle or not); it renders the effective potential of the holonomy (in principle) calculable.}
 The discrete gauge symmetry~\eqref{discshift} arises very naturally in this scenario as the four-dimensional counterpart of large gauge transformations in five dimensions. 

From this point of view, a non-compact scalar field in four dimensions could be viewed as associated to an $\mathbb{R}$ gauge theory in five dimensions, rather than to a $U(1)$ gauge theory. However there are very strong arguments against having $\mathbb{R}$ gauge theories~\cite{Banks:2010zn}. Therefore, for the non-compact scalar field, this mechanism that protects against harmful Planck-induced operators would not be applicable. It would be nice if one could find an alternative mechanism that protects against Planck-induced operators.

\begin{itemize}

\item[(ii)] Suppose the back-reaction sector produces a potential which is periodic with some scale $f$, which is smaller than the fundamental periodicity scale $f_{\rm UV}\,,$ 
\be
 f_{\rm UV}=n f > f ~.
 \ee
One should now demand that $f_{\rm UV}> \Lambda/g$. This gives
\be
\Lambda \lesssim 4\pi\, v \,n^{1\over 4}\,.\label{ns}
\ee
\end{itemize}

Unlike the previous scenario, it is actually easy to realize $f_{\rm UV}=n f > f$. For example, consider an axion field $\phi$ coupled to two strongly coupled sectors,
\be
\frac{\phi}{f_{\rm UV}} G_1 \wedge G_1+\frac{n \phi}{f_{\rm UV}} G_2 \wedge G_2
\label{db}
\ee
where $G_1$ and $G_2$ are the gauge bosons of the two sectors.  Due to non perturbative effects,  the first term above would generate a rolling potential for $\phi$, whereas the second term would  be responsible for the backreaction potential. 
Morally speaking, this amounts to having the particles in the confining sector carry charge in units of $n$ under the Peccei-Quinn symmetry. 
The drawback is that $n$ appears with the power $1/4$ in~\eq{ns}. Thus, very large values of $n$ are required in order to significantly raise the cutoff and a QFT UV-completion seems challenging. We consider two possibilities here.  First, let us imagine a UV-completion inspired by the KSVZ axion model, where  vector-like quarks are introduced to induce the terms in \eq{db},
\be
(\Phi \bar{Q}_{L1}Q_{R1}+{\rm h.c.})+(\frac{\Phi^n}{\Lambda^{n-1}} \bar{Q}_{L2}Q_{R2}+{\rm h.c.}) \, .
\ee
Here the axion field $\phi$ is the angular part of $\Phi= \rho \exp(i\phi /f_{\rm UV} )$, $\langle \rho \rangle \sim f_{\rm UV}$, $Q_{L1,R1}$ are charged under the first strong sector, and  $Q_{L2,R2}$ are charged under the strong group responsible for the back-reaction.  The vector-like quark $Q_2$ would obtain a mass $f_{\rm UV} (f_{\rm UV} / \Lambda)^{n-1}$, which is extremely small for large $n$ values (unless $(f_{\rm UV}/\Lambda)$ were tuned to be close to unity  with very high accuracy). This would lead to an effectively vanishing mass for $Q_2$ and thus to a negligible feedback potential in the back-reacting sector (as the axion potential is proportional to the lightest quark mass). Another possible way to induce the second term in \eq{db} is by having $n$ quarks in the back-reacting sector. This is, however, also problematic because a strongly coupled theory with so many flavors would not be asymptotically free, unless the number of colors were also of the same order. A very large number of colors could lead to large measurable indirect effects  in the SM sector mediated through the Higgs.

 It is possible to take more reasonable values for $n$ and raise the cut-off by a factor of 2 to 3. We use an analog of this mechanism in our explicit model in the next section.

\begin{itemize}
\item[(iii)] One could imagine having several pNGBs. One can even mimic a non-compact field by choosing a combination with incommensurate coefficients. 
\end{itemize}

This idea would not allow to increase the cutoff unless, perhaps, there is also a hierarchy among the decay constants of the different pNGBs. Suppose there are two pNGBs and the rolling potential~\eqref{Vphi} is governed by the larger decay constant of the two while the backreaction potential~\eqref{axpot} is governed by the smaller decay constant. In this extreme case there is no minimum with the desirable properties since the fields are orthogonal. In the opposite situation one does not gain anything regarding the fine-tuning problem. Also other variations of this scenario which we have considered do not produce viable vacua.  Perhaps one can sufficiently complicate this scenario in order to achieve a phenomenologically interesting model.   
An analogous string-theoretic mechanism that includes large non-periodic field excursions has been suggested in Ref.~\cite{McAllister:2008hb}.

\begin{itemize}
\item[(iv)] In principle, not all the relaxion couplings have to be given by the same $g$ and we can also imagine that the linear terms in the relaxion are absent. For example, instead of \eq{mu2} and \eq{Vphi}, we could consider
\be \label{Vphi2}
\mu^2=g_h^2 \phi^2-\Lambda^2 \, , \ \ \ 
V(\phi)=g_\phi^2 \Lambda^2 \frac{\phi^2}{2} + \cdots\,.
\ee
Here $g_\phi$ could be generated from $g_h$ by closing the Higgs loop.  
\end{itemize}

The rolling stops roughly at $\phi\sim\Lambda/g_h$, and requiring the first derivative of the potential to vanish, as in \eq{gaxion}, we get $\frac{g^2_\phi}{g_h}\sim  \frac{M^4}{\Lambda^3 f}$, with $M\lesssim 4\pi v$.
The requirement that $f \gtrsim \Lambda/g_h$ now translates into
\be
\Lambda\lesssim  \frac{M}{\sqrt{r}}
\ee
with $r \equiv g_\phi/g_h$. If it were natural to take $r\ll 1$ this would provide a way out of the no-go ``theorem''. Unfortunately, as soon as the coupling $g_h$ is present one can close the Higgs loop and generate a $\phi^2$ term in the potential $V(\phi)$ of the order $\frac{1}{16 \pi^2} \Lambda^2g_h^2\phi^2$. Therefore $r$ cannot be too small without introducing the same sort of fine tuning we have set out to eliminate. $r$ can naturally be as small as $1/(4 \pi)$, thanks to the loop factor, which can push the cut-off to the few TeV scale, 
\be
\Lambda\lesssim  10\,{\rm TeV}  \left( \frac{M}{3~{\rm TeV}} \right) \left( \frac{r}{1/4\pi} \right)^{-1/2}~ . 
\ee

We conclude that if one wants to push the cutoff well above the TeV scale, the relaxion cannot be an axion, or any other pNGB, with period $2 \pi f$.  We have presented several straightforward ways to try and avoid this conclusion and it would be interesting to see if any of them (or some other idea) could work. One can still compare the relaxion scenario to other approaches towards solving the little hierarchy problem where the relevant degrees of freedom of the effective theory are in the few-TeV range or even below (see e.g. Refs.~\cite{Chacko:2005pe, Panico:2015jxa,Bellazzini:2014yua,Schmaltz:2005ky} and references therein).

In the following section we introduce a simple concrete realization of the relaxation framework,  with the aim of elucidating the theoretical difficulties and the partial solutions mentioned above in a more concrete way. In our specific construction we will incorporate the items (ii) and (iv) in an attempt to push the cutoff to the few-TeV scale.

\section{A Familon Model} \label{sec:model}

In this section we present a calculable realization of the cosmological relaxation framework~\cite{Graham:2015cka}. In our model the rolling field $\phi$ is a familon, the pNGB  of a spontaneously broken flavor symmetry.  We use this model to demonstrate explicitly the points of the previous section and we also analyze its phenomenological properties.

We will assume that the period  $f_{\rm UV}$ of the rolling field is related to the period $f$ that appears in the low energy effective action as 
\be\label{effper}
f_{\rm UV} = 2nf\, ,\qquad n\in \mathbb{N}~.
 \ee
This can be achieved by assuming that the fields in the back-reacting sector carry charges in integer units of $n$. (The origin of the factor of 2 in~\eqref{effper} will be clear below.) 
Our Lagrangian for the back-reacting sector is
\be \label{Lfamilon}
{\cal L}= -y_1  e^{i \frac{2n \phi}{f_{\rm UV}}} \epsilon^{\alpha \beta} h_\alpha L_\beta N - y_2 h^{\dagger \alpha} L_\alpha^c  N - m_L  \epsilon^{\alpha \beta} L_\alpha L_\beta^c  - \frac{m_N}{2} N N + {\rm h.c.}
\,.\ee
We use two-component spinor notation for the fermions, $\epsilon^{\alpha \beta}$ is the antisymmetric symbol of $SU(2)_L$, $h^{T \alpha} = (h^+, h^{0} )$. $L$ and $L^c$ are doublets under the $SU(2)_L$ gauge group of the SM, with opposite hypercharge,
\be
L_\alpha=\begin{pmatrix}
\nu\\
E\\
\end{pmatrix}
\qquad
L_\alpha^c=\begin{pmatrix}
E^c\\
\nu^c\\
\end{pmatrix} \, ,
\ee
$N$ is a SM singlet, $\phi$ is the familon field. (It can be realized, for instance, when a flavon field $\Phi = \rho \exp[i\phi/f_{\rm UV}]$ aquires a VEV.)

If $m_N=0$ (but we always keep $m_L\neq 0$) the model has a $U(1)_{NL}$ global symmetry under which the fields transform as follows:

\begin{center}
  \begin{tabular}{ l   | r }
    \hline
     &  $U(1)_{NL}$ \\ \hline
    $N$ & -n \\ \hline
    $L$  & -n \\  \hline
    $L^c$ & n \\  \hline
   $h$ & 0 \\  \hline
   $e^{i\phi/f_{\rm UV}}$ &   1 \\  \hline
   SM  & 0 \\  \hline
  \end{tabular}
\end{center}
 The normalization of the charges under $U(1)_{NL}$ is chosen so that the flavon field has unit charge. Note that an effective periodicity as in~\eqref{effper} appears in the back-reacting sector.

Clearly,  $m_N\neq 0$ explicitly (softly) breaks the $U(1)_{NL}$ symmetry. For $m_N=0$, the continuous shift symmetry prevents any potential for $\phi$. 

For $m_N\neq 0$ a two-loop potential for the familon is generated even in the Electroweak-preserving vacuum.
We first describe the one-loop analysis where a potential for $\phi$ is generated once the Higgs gets a VEV.
 We will then come back to the issue of two-loop corrections.

The one-loop Coleman-Weinberg potential for $\phi$ is
\be \label{CWdef}
V_{\rm CW}(\phi) = -\frac{\Lambda^2}{16 \pi^2} {\rm Tr} \left[M^\dagger (\phi) M(\phi) \right] -\frac{1}{32 \pi^2} {\rm Tr} \left[ \left( M^\dagger (\phi) M(\phi) \right)^2 \log \frac{M^\dagger (\phi) M(\phi)}{\Lambda^2} \right] \, ,
\ee
where the mass matrix for the fermions in the $\{ N,\nu,\nu^c \}$ basis is 
\be \label{massmatrix}
M(\phi) = 
\begin{pmatrix}
m_N& - y_1 h^0 U & y_2 h^{0*}  \\
- y_1 h^0 U & 0 & m_L\\
y_2 h^{0*}  & m_L & 0
\end{pmatrix} \,,
\ee
with $U\equiv e^{i\frac{2n\phi}{f_{\rm UV}}}\,.$
The term relevant to our discussion is given by 
\be \label{VCWbefore}
V_{\rm CW} (\phi) \simeq  -\frac{1}{4 \pi^2} m_L m_N  y_1 y_2 |h^0|^2  \cos \left(\frac{\phi}{f}\right)  \log\left( \frac{\Lambda^2}{\tilde{m}^2}\right)\,,
\ee
where $\tilde{m}$ is the larger of $m_L$ and $m_N$. Upon EWSB, $\langle h^0 \rangle = v = 174$ GeV, this gives the contribution
\be \label{VCWafter}
V^{\rm EWSB}_{\rm CW}(\phi)= -\frac{1}{4 \pi^2} m_L m_N  y_1 y_2 v^2\cos \left(\frac{\phi}{f}\right)\log\left( \frac{\Lambda^2}{\tilde{m}^2}\right)\,.
\ee
From this potential we can find the mass of $\phi$ by expanding around the minimum:
\be \label{mphi}
m_\phi \simeq 5 \ {\rm GeV} \left(\frac{m_L}{900 {\rm~GeV}}\right)^\frac{1}{2} \left(\frac{m_N}{900 {\rm~GeV}}\right)^\frac{1}{2}  \left(\frac{y_1}{1}\right)^\frac{1}{2} \left(\frac{y_2}{1}\right)^\frac{1}{2} \left( \frac{10 \ {\rm TeV}}{f} \right) \, . 
\ee

As we mentioned above, the symmetries allow for the generation of a potential for $\phi$ even before electroweak symmetry breaking.\footnote{For a related discussion see Ref.~\cite{Espinosa:2015eda}.} 
Such a contribution takes the form
\be
V^{\rm 2-loop}_{\rm CW}(\phi)\sim- \frac{1}{4 \pi^2} m_L m_N  y_1 y_2 \left(\frac{\Lambda_c^2}{16 \pi^2}\right) \cos \left(\frac{\phi}{f}\right)\,.
\label{2loop}
\ee
One can think of this as coming from~\eqref{VCWbefore} by contracting $h^0$ and $h^{0*}$ in an additional loop. (In~\eqref{2loop} we have suppressed the logarithm.)
$\Lambda_c$ is the scale at which the Higgs loop is cut off. In order for $\phi$ not to stop rolling before EWSB we must require $V^{\rm 2-loop}_{\rm CW}(\phi) < V^{\rm EWSB}_{\rm CW}(\phi)$. This gives the condition  
\be
\Lambda_c\lesssim 4 \pi v \, .
\label{relmir}
\ee

Eq.~\eqref{relmir} can be satisfied in a simple  extension of our model where $m_{N}$ is generated via a mini-See-Saw mechanism from the following Lagrangian
\be \label{Lfamilon2}
{\cal L}=  -y_1  U \epsilon^{\alpha \beta} h_\alpha L_\beta N - y_2 h^{\dagger \alpha} L_\alpha^c  N - m_L  \epsilon^{\alpha \beta} L_\alpha L_\beta^c  - m_D N N^c-\frac{m_{N^c}}{2} N^c N^c+ {\rm h.c.\,.}
\ee
We have just added to~\eqref{Lfamilon} a new fermion $N^c$ with a Majorana mass $m_{N^c} \ll m_D$. If we integrate out $N^c$ we obtain the original model~\eqref{Lfamilon}
with $m_N\sim{m^2_D}/{m_{N^c}}\,$. 
$V(\phi)$ has no quadratically divergent contributions from momenta larger than $m_{N^c}$. To prove this we first observe that the quadratically divergent piece must be analytic in the couplings. Second, we observe that if we set $m_L=0$ the Coleman-Weinberg potential for $U$ must vanish as the $U(1)_{NL}$ would be restored. (The argument is the same as in the original model~\eqref{Lfamilon}.) Similarly, if we set $m_D=0$ or $m_{N^c}=0$ the Coleman-Weinberg potential must vanish. Therefore the two-loop potential for $U$ must be of the form $(y_1  y_2  m_L m_D^2 m_{N^c}U +{\rm h.c.\,})$, which means that in the extended model~\eqref{Lfamilon2} there can be  a log divergence at most. Taking $m_{N^c}$ large compared to the other masses, this expression reduces to~\eqref{2loop} with $\Lambda_c\sim m_{N^c}$. \eq{relmir} then implies an upper bound on $m_{N^c}$ of roughly 3 TeV. This guarantees that as long as $m_{N^c} \lesssim 3$ TeV the two-loop effects~\eqref{2loop} are small and do not spoil the relaxation mechanism.

 Suppose that, in addition to the back-reacting sector described above,  we  add a heavier  sector  (for instance a sector  very similar to \eq{Lfamilon2} but with $U(1)_{NL}$ charges of order $\mathcal{O}(1)$ and masses at the scale $4\pi M$)  which explicitly breaks the shift symmetry and generates the terms
  \bea \label{heavy}
V(h) &=& \left[- \Lambda^2 +M^2 \cos \left( \frac{\phi}{f_{\rm UV}} \right)\right]h^\dagger h+\lambda (h^\dagger h)^2~, \nonumber\\
V(\phi)&=& \frac{\Lambda^2 M^2}{16 \pi^2} \cos \left( \frac{\phi}{f_{\rm UV}} \right) \, .
 \eea
These provide us with the analogs of  \eqref{Higgspotential}, \eqref{mu2}, \eqref{Vphi}. The periodicity in this sector is $\phi \to \phi + 2\pi  f_{\rm UV}$, corresponding to the fundamental gauge symmetry. 
 Since we are imagining that the origin of~\eqref{heavy} is from a sector of heavy fields, $V(h)$ arises at one loop while $V(\phi)$ is generated only at two loops. We take $M$ and $\Lambda$ to be of the same order, but require $M > \Lambda$ in order for the Higgs to condense.

Now we are ready to work out the phenomenological implications of our model. The familon $\phi$ stops rolling when $\frac{\partial}{\partial \phi}[V(\phi) + V_{\rm CW}^{\rm EWSB}(\phi)]= 0$\,, which gives 
\be
\Lambda \sim    \left[4 m_L m_N y_1 y_2 v^2 \frac{f_{\rm UV}}{f}\,\log\left({m^2_{N^c}\over \tilde{m}^2}\right)\right]^\frac{1}{4} \,,\label{derivnew}
\ee
after using
 \be
 \mu^2\sim 0 \Rightarrow M^2 \sim{\Lambda^2\over \cos\left({\phi\over f}\right)}\,.
 \ee
 This places the cutoff at
 \bea
\Lambda &\sim& 5{\rm~TeV} \left(\frac{m_L}{900 {\rm~GeV}}\right)^\frac{1}{4} \left(\frac{m_N}{900 {\rm~GeV}}\right)^\frac{1}{4}  \left(\frac{y_1}{4 \pi }\right)^\frac{1}{4} \left(\frac{y_2}{4 \pi}\right)^\frac{1}{4}  \left( \frac{n}{10} \right)^\frac{1}{4}\,.\label{derivnew}
\eea
As we anticipated in the previous section, $\Lambda$ is of order TeV unless $f_{\rm UV}\gg f$, which would require the fundamental charge $n$ in the back-reacting sector to be very large. A very large $n$, besides being aesthetically unappealing, is difficult to obtain in UV completions of this model. The reason is that for $n \gg 1$
the Yukawa coupling $y_1$ would generally  arise from an extremely irrelevant operator, making the back-reaction negligible.

\begin{figure}[!t]
\centering
\includegraphics[width=0.8 \textwidth]{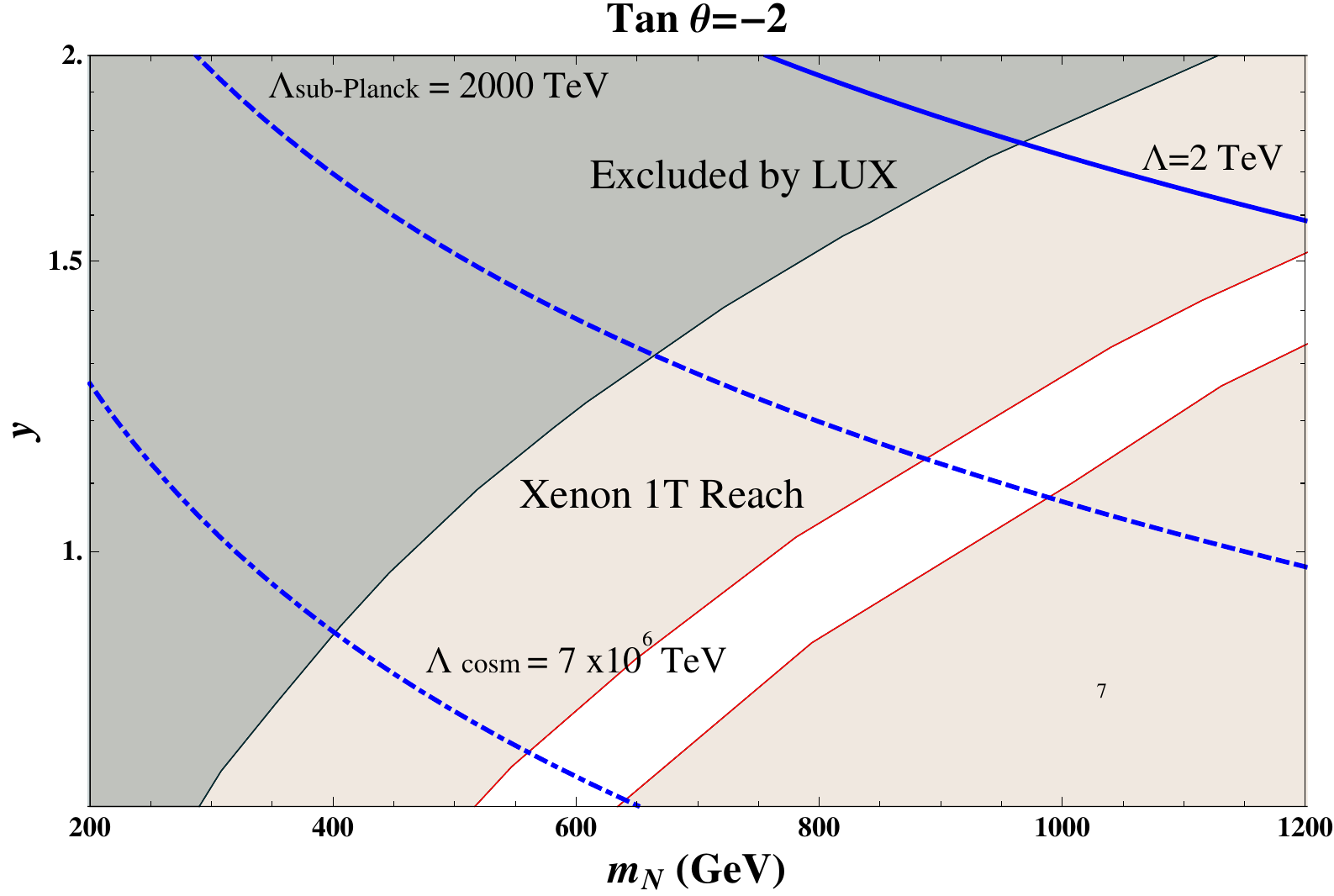}
{\small \caption{\small  \label{fig}
 Constraints from dark matter phenomenology on our parameter space. We define $y_1=y \cos \theta$, $y_2=y \sin \theta$\,, and fix $\tan \theta = -2$. The parameter $m_L$ is adjusted at each point to match the observed relic density. The region shaded in grey is excluded by LUX, while the brown region will be probed in the near future by the XENON 1\,Ton experiment. The blue lines correspond to the following cutoffs: the one saturating the cosmological constraints of Eq.~\eqref{eq:cutoff} with $f= \frac{f_{\rm UV}}{2 n} = 10$ TeV (dot-dashed), the one from Eq.~\eqref{derivnew} with $f_{\rm UV} = M_{{\rm Pl}}$, so that the field excursion remains sub-Planckian, and $f= 10$ TeV (dashed), and  the one from Eq.~\eqref{derivnew} with $n=10$ (solid). }}
\end{figure}

In Ref.~\cite{Graham:2015cka} it has been shown that cosmological requirements enforce the inequalities
 \be \label{venergy}
 \frac{\Lambda^2}{M_{\rm Pl}}<H<(V'(\phi))^\frac{1}{3}\,.
\ee
The first inequality comes from requiring that the energy density of $\phi $ ($\sim \Lambda^4$) be lower than the energy density of the inflaton. The second arises from asking that the classical rolling dominate over the quantum fluctuations.
In our scenario the condition of \eqref{venergy} is satisfied as long as
\begin{align}
\Lambda & < M_{\rm Pl}^\frac{1}{2} \left[ \frac{m_L m_N y_1 y_2 v^2}{2 \pi^2 f_{\rm UV}} n \right]^\frac{1}{6} \nonumber \\
& \simeq  10^7{\rm~TeV} \left(\frac{m_L}{900 {\rm~GeV}}\right)^\frac{1}{6} \left(\frac{m_N}{900 {\rm~GeV}}\right)^\frac{1}{6}  \left(\frac{y_1}{4 \pi }\right)^\frac{1}{6} \left(\frac{y_2}{4 \pi}\right)^\frac{1}{6}  \left( \frac{n}{10} \right)^\frac{1}{6} \left(\frac{f_{\rm UV}}{100 {\rm~TeV}}\right)^{-\frac{1}{6}}  \label{eq:cutoff} \, . 
\end{align}
This is obviously satisfied given \eqref{derivnew}.\footnote{Note that \eqref{derivnew} scales as $n^{1/4}$, while \eqref{eq:cutoff} scales as $n^{1/6}$. This implies that $n$ cannot be arbitrarily large otherwise we would not comply with the cosmological requirements of the relaxation mechanism. In any case, as already mentioned, values of  $n \gg 1$ are very difficult to achieve in UV completions of  this model and as a result it seems difficult to obtain a cutoff higher than what quoted in \eqref{derivnew}.}

We now  briefly comment on the dark-matter and collider aspects of the model~\eqref{Lfamilon2}. 
The dark matter candidate is a Majorana fermion corresponding to the lightest eigenstate of the mass matrix \eqref{massmatrix}.
 Constraints on the parameter space of this model coming from relic abundance, direct detection, indirect detection and collider searches have been studied in detail in previous literature~\cite{Cohen:2011ec, Cheung:2013dua, Calibbi:2015nha}. Here, we use the results of Ref.~\cite{Cheung:2013dua} to summarize the most important constraints in Fig.~\ref{fig}. In the plot, $m_L$ is adjusted at each point by requiring the right relic density of dark matter and we define $\theta$ by $y_1 = y \cos \theta$ and $y_2=y \sin \theta$. The region shaded in grey is excluded by LUX, while the brown region will be probed in the near future by the XENON 1 Ton.  We also show contours for three values of the cutoff. The maximal value satisfying the constraints from cosmology of~\eqref{eq:cutoff} is shown in dotted-dashed-blue; the value which fulfils the requirement to have a sub-Planckian $f_{\rm UV}$ is shown in dashed-blue; the value from \eqref{derivnew} with $n = 10$ is shown in solid-blue. 

The collider phenomenology of this framework differs from what is typically predicted by other models that address the little hierarchy problem. 
Namely, there are no light top-partner fields or partners of the electroweak gauge bosons. Instead, there  are vector-like fermions, with at least one of them being an $SU(2)_L$ doublet. 
For the models not to be fine-tuned, these new degrees of freedom should be present at or below the TeV scale.

\bigskip

\appendix
\begin{center}
\large{\textbf{Acknowledgements}}
\end{center}
We are grateful to  Prateek Agrawal, Kfir Blum, Clliford Cheung, Lorenzo di Pietro, Diptimoy Ghosh, Roni Harnik, Alex Pomarol, Pedro Schwaller, Emmanuel Stamou, and Tomer Volansky  for very insightful  discussions.
GP is supported by the BSF (2014230), IRG (249185), ISF (687/14), and ERC-2013-CoG (614794) grants. ZK is supported in part by an Israel Science Foundation center for excellence grant, by the I-CORE program of the Planning and Budgeting Committee and the
ISF (grant number 1937/12), by the ERC STG grant 335182, and by the BSF under grant 2010/629.

\newpage

\bibliographystyle{JHEP}
\bibliography{Familon}

\providecommand{\href}[2]{#2}\begingroup\raggedright\begin{thebibliography}{10}

\bibitem{Graham:2015cka}
P.~W. Graham, D.~E. Kaplan, and S.~Rajendran, {\it {Cosmological Relaxation of
  the Electroweak Scale}},  \href{http://arxiv.org/abs/1504.07551}{{\tt
  arXiv:1504.07551}}.

\bibitem{Dvali:2003br}
G.~Dvali and A.~Vilenkin, {\it {Cosmic attractors and gauge hierarchy}},  {\em
  Phys. Rev.} {\bf D70} (2004) 063501,
  [\href{http://arxiv.org/abs/hep-th/0304043}{{\tt hep-th/0304043}}].

\bibitem{Espinosa:2015eda}
J.~R. Espinosa, C.~Grojean, G.~Panico, A.~Pomarol, O.~Pujol\`as, and
  G.~Servant, {\it {Cosmological Higgs-Axion Interplay for a Naturally Small
  Electroweak Scale}},  \href{http://arxiv.org/abs/1506.09217}{{\tt
  arXiv:1506.09217}}.

\bibitem{Hardy:2015laa}
E.~Hardy, {\it {Electroweak relaxation from finite temperature}},
  \href{http://arxiv.org/abs/1507.07525}{{\tt arXiv:1507.07525}}.

\bibitem{Patil:2015oxa}
S.~P. Patil and P.~Schwaller, {\it {Relaxing the Electroweak Scale: the Role of
  Broken dS Symmetry}},  \href{http://arxiv.org/abs/1507.08649}{{\tt
  arXiv:1507.08649}}.

\bibitem{Antipin:2015jia}
O.~Antipin and M.~Redi, {\it {The Half-composite Two Higgs Doublet Model and
  the Relaxion}},  \href{http://arxiv.org/abs/1508.01112}{{\tt
  arXiv:1508.01112}}.

\bibitem{Jaeckel:2015txa}
J.~Jaeckel, V.~M. Mehta, and L.~T. Witkowski, {\it {Musings on cosmological
  relaxation and the hierarchy problem}},
  \href{http://arxiv.org/abs/1508.03321}{{\tt arXiv:1508.03321}}.

\bibitem{Kilic:2015joa}
C.~Kilic and S.~Swaminathan, {\it {Can A Pseudo-Nambu-Goldstone Higgs Lead To
  Symmetry Non-Restoration?}},  \href{http://arxiv.org/abs/1508.05121}{{\tt
  arXiv:1508.05121}}.

\bibitem{Coleman:1969sm}
S.~R. Coleman, J.~Wess, and B.~Zumino, {\it {Structure of phenomenological
  Lagrangians. 1.}},  {\em Phys. Rev.} {\bf 177} (1969) 2239--2247.

\bibitem{Wess:1992cp}
J.~Wess and J.~Bagger, {\em {Supersymmetry and supergravity}}.
\newblock 1992.

\bibitem{Ray:2006wk}
S.~Ray, {\it {Some properties of meta-stable supersymmetry-breaking vacua in
  Wess-Zumino models}},  {\em Phys. Lett.} {\bf B642} (2006) 137--141,
  [\href{http://arxiv.org/abs/hep-th/0607172}{{\tt hep-th/0607172}}].

\bibitem{Sun:2008nh}
Z.~Sun, {\it {Continuous degeneracy of non-supersymmetric vacua}},  {\em Nucl.
  Phys.} {\bf B815} (2009) 240--255,
  [\href{http://arxiv.org/abs/0807.4000}{{\tt arXiv:0807.4000}}].

\bibitem{Komargodski:2009jf}
Z.~Komargodski and D.~Shih, {\it {Notes on SUSY and R-Symmetry Breaking in
  Wess-Zumino Models}},  {\em JHEP} {\bf 04} (2009) 093,
  [\href{http://arxiv.org/abs/0902.0030}{{\tt arXiv:0902.0030}}].

\bibitem{Curtin:2012yu}
D.~Curtin, Z.~Komargodski, D.~Shih, and Y.~Tsai, {\it {Spontaneous R-symmetry
  Breaking with Multiple Pseudomoduli}},  {\em Phys. Rev.} {\bf D85} (2012)
  125031, [\href{http://arxiv.org/abs/1202.5331}{{\tt arXiv:1202.5331}}].

\bibitem{Affleck:1983mk}
I.~Affleck, M.~Dine, and N.~Seiberg, {\it {Dynamical Supersymmetry Breaking in
  Supersymmetric QCD}},  {\em Nucl. Phys.} {\bf B241} (1984) 493--534.

\bibitem{Affleck:1984xz}
I.~Affleck, M.~Dine, and N.~Seiberg, {\it {Dynamical Supersymmetry Breaking in
  Four-Dimensions and Its Phenomenological Implications}},  {\em Nucl. Phys.}
  {\bf B256} (1985) 557.

\bibitem{Dumitrescu:2010ca}
T.~T. Dumitrescu, Z.~Komargodski, and M.~Sudano, {\it {Global Symmetries and
  D-Terms in Supersymmetric Field Theories}},  {\em JHEP} {\bf 11} (2010) 052,
  [\href{http://arxiv.org/abs/1007.5352}{{\tt arXiv:1007.5352}}].

\bibitem{Banks:2010zn}
T.~Banks and N.~Seiberg, {\it {Symmetries and Strings in Field Theory and
  Gravity}},  {\em Phys. Rev.} {\bf D83} (2011) 084019,
  [\href{http://arxiv.org/abs/1011.5120}{{\tt arXiv:1011.5120}}].

\bibitem{ArkaniHamed:2003wu}
N.~Arkani-Hamed, H.-C. Cheng, P.~Creminelli, and L.~Randall, {\it {Extra
  natural inflation}},  {\em Phys. Rev. Lett.} {\bf 90} (2003) 221302,
  [\href{http://arxiv.org/abs/hep-th/0301218}{{\tt hep-th/0301218}}].

\bibitem{Hosotani:1983xw}
Y.~Hosotani, {\it {Dynamical Mass Generation by Compact Extra Dimensions}},
  {\em Phys. Lett.} {\bf B126} (1983) 309.

\bibitem{Rattazzi:2000hs}
R.~Rattazzi and A.~Zaffaroni, {\it {Comments on the holographic picture of the
  Randall-Sundrum model}},  {\em JHEP} {\bf 04} (2001) 021,
  [\href{http://arxiv.org/abs/hep-th/0012248}{{\tt hep-th/0012248}}].

\bibitem{Feng:2003mk}
B.~Feng, M.-z. Li, R.-J. Zhang, and X.-m. Zhang, {\it {An inflation model with
  large variations in spectral index}},  {\em Phys. Rev.} {\bf D68} (2003)
  103511, [\href{http://arxiv.org/abs/astro-ph/0302479}{{\tt
  astro-ph/0302479}}].

\bibitem{Cacciapaglia:2007fw}
G.~Cacciapaglia, C.~Csaki, J.~Galloway, G.~Marandella, J.~Terning, and
  A.~Weiler, {\it {A GIM Mechanism from Extra Dimensions}},  {\em JHEP} {\bf
  04} (2008) 006, [\href{http://arxiv.org/abs/0709.1714}{{\tt
  arXiv:0709.1714}}].

\bibitem{Perez:2008ee}
G.~Perez and L.~Randall, {\it {Natural Neutrino Masses and Mixings from Warped
  Geometry}},  {\em JHEP} {\bf 01} (2009) 077,
  [\href{http://arxiv.org/abs/0805.4652}{{\tt arXiv:0805.4652}}].

\bibitem{McAllister:2008hb}
L.~McAllister, E.~Silverstein, and A.~Westphal, {\it {Gravity Waves and Linear
  Inflation from Axion Monodromy}},  {\em Phys. Rev.} {\bf D82} (2010) 046003,
  [\href{http://arxiv.org/abs/0808.0706}{{\tt arXiv:0808.0706}}].

\bibitem{Chacko:2005pe}
Z.~Chacko, H.-S. Goh, and R.~Harnik, {\it {The Twin Higgs: Natural electroweak
  breaking from mirror symmetry}},  {\em Phys. Rev. Lett.} {\bf 96} (2006)
  231802, [\href{http://arxiv.org/abs/hep-ph/0506256}{{\tt hep-ph/0506256}}].

\bibitem{Panico:2015jxa}
G.~Panico and A.~Wulzer, {\it {The Composite Nambu-Goldstone Higgs}},
  \href{http://arxiv.org/abs/1506.01961}{{\tt arXiv:1506.01961}}.

\bibitem{Bellazzini:2014yua}
B.~Bellazzini, C.~Csaki, and J.~Serra, {\it {Composite Higgses}},  {\em Eur.
  Phys. J.} {\bf C74} (2014), no.~5 2766,
  [\href{http://arxiv.org/abs/1401.2457}{{\tt arXiv:1401.2457}}].

\bibitem{Schmaltz:2005ky}
M.~Schmaltz and D.~Tucker-Smith, {\it {Little Higgs review}},  {\em Ann. Rev.
  Nucl. Part. Sci.} {\bf 55} (2005) 229--270,
  [\href{http://arxiv.org/abs/hep-ph/0502182}{{\tt hep-ph/0502182}}].

\bibitem{Cohen:2011ec}
T.~Cohen, J.~Kearney, A.~Pierce, and D.~Tucker-Smith, {\it {Singlet-Doublet
  Dark Matter}},  {\em Phys. Rev.} {\bf D85} (2012) 075003,
  [\href{http://arxiv.org/abs/1109.2604}{{\tt arXiv:1109.2604}}].

\bibitem{Cheung:2013dua}
C.~Cheung and D.~Sanford, {\it {Simplified Models of Mixed Dark Matter}},  {\em
  JCAP} {\bf 1402} (2014) 011, [\href{http://arxiv.org/abs/1311.5896}{{\tt
  arXiv:1311.5896}}].

\bibitem{Calibbi:2015nha}
L.~Calibbi, A.~Mariotti, and P.~Tziveloglou, {\it {Singlet-Doublet Model: Dark
  matter searches and LHC constraints}},
  \href{http://arxiv.org/abs/1505.03867}{{\tt arXiv:1505.03867}}.

\end{thebibliography}\endgroup

\end{document}